# MOSAIC: MOBILE OBJECT SEGMENTATION UNDER ADVERSE IMAGING CONDITIONS FOR RAPID L-PBF KEYHOLE BEHAVIOR CHARACTERIZATION

**Garrett Mathesen**
Northwestern University
Evanston, IL

**Kyle Mumm**
Northwestern University
Evanston, IL

**Carter Taylor**
Northwestern University
Evanston, IL

**Jian Cao**
Northwestern University
Evanston, IL

## ABSTRACT

*In laser powder bed fusion (L-PBF) processes, the rapid evolution of gas and fluid interactions complicates our ability to properly monitor or control the process, with unstable keyholes leading to porosity and spatter formation. High-speed operando x-ray imaging of the keyhole has been used to better understand the impact of these interactions on the monitoring and control of the L-PBF process. MOSAIC, a Mobile Object Segmentation algorithm for experiments under Adverse Imaging Conditions, is designed to perform rapid analysis of keyhole dynamics during active beamline experimentation without needing time-consuming manual labeling or model training. Validation studies performed on 12 unique samples proved the robustness of MOSAIC with an average F1 score of 0.894 and a precision of 0.953 when compared to manually segmented images, performing equally or better than the SAM and YOLO machine learning methods tested. MOSAIC is efficient, processing frames cropped to a moving window approximately 150x250 pixels at 19.9 milliseconds per image on CPU, compared to 54 and 5284 milliseconds per image for inference on CPU for YOLO and SAM models.*

## 1. INTRODUCTION

The use of synchrotron x-ray sources for high-speed operando imaging of the Laser Powder Bed Fusion (L-PBF) additive manufacturing process has enabled novel insights vital for process control. By revealing complex dynamics of the keyhole during the L-PBF process, phenomena such as defect formation, melt pool fluid flow, and the effective absorptivity of the laser can be studied quantitatively. While numerous papers have been published on observations from high-speed operando x-ray imaging of the L-PBF process, a current limiting factor for researchers is the time-consuming task of segmenting the keyhole from the recorded images. We propose a Mobile Object Segmentation under Adverse Imaging Conditions (MOSAIC) algorithm to perform precise keyhole segmentation without requiring a fine-tuned model or complex computational hardware to run efficiently. With this method, those performing L-PBF experiments at x-ray imaging beamlines will be able to improve experimental efficiency and quality through informed adaptation of experimental parameters on-site.

Segmentation of the keyhole plays a critical role in ensuring well-informed process monitoring approaches are used when controlling L-PBF processes. In-situ monitoring and control is rapidly developing into an adequate method of defect detection or prevention. Many of the proposed control methods rely on high-fidelity simulations to optimize parameters [1,2], data-based models that attempt to abstract the underlying physics [3–5], or some combination of both [6,7]. While authors have successfully explored in-line feedback control schemes to maintain melt pool temperatures [8], the reliance on trained machine learning (ML) models or high fidelity models presents a challenge when transferring to industrial applications. The large amount of training data needed for ML models and long simulation compute time may not be suitable for a wide range of machine designs, sensor setups, data formats, and the quick turnaround time demanded by industrial L-PBF manufacturers. The speed and robustness requirements of sensing capability for in-situ process control emphasize the need for developing quicker physics-based models or other forms of surrogate modeling, especially for the chaotic dynamics of the keyhole during high-energy process regimes. To validate sensor predictions, a ground truth must be created for the keyhole dynamics and morphology, which is most accurately generated from high-speed operando imaging of the L-PBF process.

Zhao et al. established one of the first L-PBF process replicators at the Argonne National Laboratory's Advanced

Photon Source (APS), using it to conduct operando imaging of the laser melting process at 50 kHz [9]. A sample figure of the L-PBF set up with respect to the incident x-ray beam can be seen in **Figure 1**. Beamline 32-ID at APS, where the process replicator is used for high-speed imaging, has a maximum effective framerate of 6.5 MHz as reported in literature [10]. A nominal framerate of 50 kHz is used in most works due to the relatively short imaging times that the MHz imaging mandates. These high-speed operando imaging experiments using the L-PBF process replicator have allowed for a number of important discoveries regarding our understanding of the L-PBF process, enabling production of keyhole stability metrics [15], direct linkages between keyhole instability and the formation of pores [16], validation of simulated powder-gas interactions using ground truth images [17], and the characterization of the vapor plume's response to keyhole characteristics [18]. The development of a robust physics understanding of the keyhole, enabled by high-speed operando imaging of the L-PBF process, leads to better defect detection models [19–22]. In all these studies, ground truth data of quantitative keyhole characteristics is required, and in many cases, the time series of keyhole morphology is needed as well.

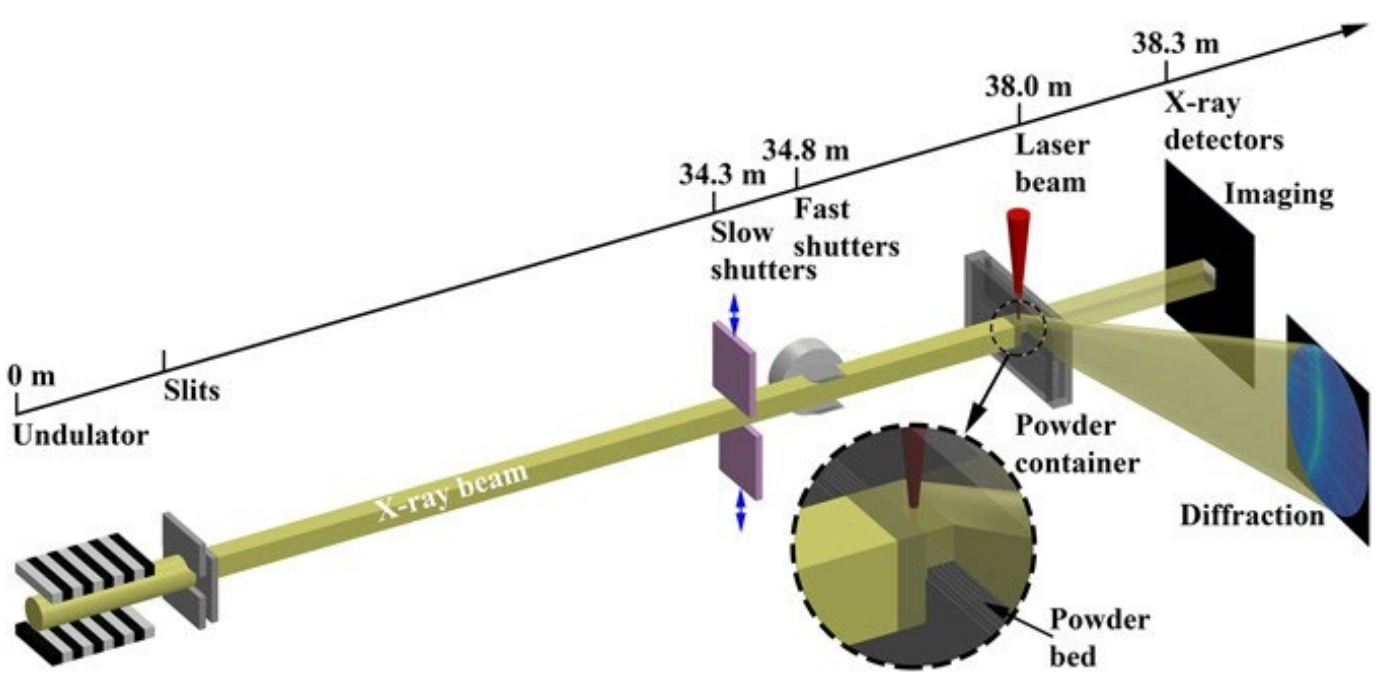

**FIGURE 1:** X-ray imaging setup from APS beamline 32-ID showing source, process replicator, and imaging detector. Reproduced from [9], licensed under CC BY 4.0.

A range of methods have been used to segment the relevant features of the keyhole morphology and interactions, which can be broadly classified as classical image processing and machine learning models. Of a selection of papers that utilize classical image processing, the main investigations performed were either determining the average behavior across the experiment [11,15,16] or performing a time-series extraction of the keyhole performed on 1-4 samples total [17,18]. When examining where ML models were used for segmentation of beamline images, the predominant mode of analysis is with databases with tens to hundreds of experiments where detailed time-series information is desired to determine single event impacts on the melt pool [19,20]. Some authors who ultimately relied on classical thresholding and image processing approaches did attempt to use ML models but noted inconsistencies in the results. This lead Huang et. al. [17] to rely on the classical image processing techniques for analysis of results where time-dependent factors were of interest.

Machine learning image segmentation tools have been used for analysis, but typically require 10,000 to 100,000 labeled images for training. Even when leveraging pretrained foundation models, domain-specific fine-tuning may still require 10,000 to 20,000 additional mask annotations, representing a substantial labeling effort [21]. The generation of training datasets for machine learning methods or performing manual labeling reduces the experimental efficiency and complicates the ability to compare results from various research groups. This challenge has motivated us to develop a deterministic image segmentation tool that can rapidly detect the boundaries of the keyhole morphology, thus giving insight into complex physical interactions. MOSAIC improves the efficiency of allocated beamtime and aims to highlight new discoveries during active experimentation, maximizing the impact of limited beamline resources. The MOSAIC algorithm can be used as a standardized analysis tool and as a ground truth for automated analysis workflows, which can drive new discoveries in the field of L-PBF monitoring and control. Additionally, the tool can be utilized in adverse imaging scenarios where the object under scientific observation has moving low-contrast boundaries that complicate common image processing methods.

## 2. MATERIALS AND METHODS

All experimental data presented in this work were collected at Argonne National Laboratory's Advanced Photon Source using beamline 32-ID. MOSAIC consists of the preprocessing pipeline (Section 2.1) and the segmentation algorithm (Section 2.2). For determining the accuracy of MOSAIC, 12 experiments with different materials, framerates, and process parameters were selected to highlight the adaptable nature of the software. **Tables 1a** and **1b** outline the relevant imaging setup details and processing parameters of the experiments performed at APS. The operando imaging setup with the process replicator has been discussed extensively in other works [9,19] and a basic configuration can be seen in **Figure 1**. For implementation of the proposed MOSAIC algorithm, Python was utilized with the help of functions from the Numpy, SciPy, and Scikit-Image libraries [22,23]. Validation studies were performed using two benchmark machine learning segmentation models: a fine-tuned You Only Look Once (YOLO) model, a deep learning object detection and segmentation framework, and the Segment Anything Model (SAM), a foundation model for promptable image segmentation. Both models were implemented and evaluated using the Ultralytics Python library [24].

**TABLE 1a:** List of samples with the relevant camera parameters.

| ID | Camera Used | Frame Rate (kHz) | Pixel Size (µm) | Shutter Speed (ns) | Frame Size |
|---|---|---|---|---|---|
| A-H | Photon FastCam SA-Z | 50 | 1.95 | 1002 | 1024 ×400 |
| I | | | | 7500 | |
| J | | | | | 896×448 |
| K, L | Shimadzu HPV-X2 | 1080 | 3.2 | 110 | 400 × 250 |

**TABLE 1b:** List of samples used for testing MOSAIC with relevant materials and process regime characteristics.

<table>
<tr><th>ID</th><th>Mat.</th><th>Power (W)</th><th>Scan (mm/s)</th><th>Powder</th><th>Substrate</th><th>Cite</th></tr>
<tr><td>A</td><td rowspan="2">AlSi10 Mg</td><td rowspan="3">200</td><td rowspan="4">1400</td><td>No</td><td rowspan="8">As-Built</td><td rowspan="8">[19]</td></tr>
<tr><td>B</td><td rowspan="3">Yes</td></tr>
<tr><td>C</td><td rowspan="2">Ti64</td></tr>
<tr><td>D</td><td>400</td></tr>
<tr><td>E</td><td rowspan="2">AlSi10 Mg</td><td rowspan="2">360</td><td rowspan="2">900</td><td>No</td></tr>
<tr><td>F</td><td rowspan="3">Yes</td></tr>
<tr><td>G</td><td rowspan="6">Ti64</td><td>150</td><td rowspan="2">800</td></tr>
<tr><td>H</td><td>350</td></tr>
<tr><td>I</td><td rowspan="4">200</td><td rowspan="3">500</td><td>No</td><td rowspan="4">Wrought</td><td rowspan="4">[25]</td></tr>
<tr><td>J</td><td>Yes</td></tr>
<tr><td>K</td><td rowspan="2">No</td></tr>
<tr><td>L</td><td>600</td></tr>
</table>

### 2.1 Preprocessing Pipeline

The MOSAIC preprocessing pipeline is designed to increase the contrast of the moving object (here, the keyhole) by performing multiple operations to reduce the noise caused by the thermal expansion of the substrate and bulk motion of the powder particles. To achieve this goal, Algorithm 1 was developed, which inputs a frame stack of raw frames $\boldsymbol{F}$ and outputs the processed image stack $\boldsymbol{P}$. Symbols used in Algorithm 1 include the frame index number $i$, such that $i \in [0, N]$, the corresponding toolpath masks $\boldsymbol{M}$, the temporal window size $k$, and the desired median filter radius $r$.

**Algorithm 1:** Preprocessing Images

**Input:** $\boldsymbol{F},\ \boldsymbol{M},\ k,\ r$

$\boldsymbol{B}(i,k) = \sum_{j=(i-k/2)}^{i+k/2} \frac{f_j}{k}$

**for** $i = 1,2,\dots,N$ **do**

$f_i = \boldsymbol{F}(i,x,y),\ b_i = \boldsymbol{B}(i,k),\ m_i = \boldsymbol{M}(i)$

$d = f_i - b_i$

$c = d[m_i]$

$c_b = \mathcal{F}^{-1}\{\mathcal{F}\{c\} \cdot G_{bw}(\mathcal{F}\{c\})\}$

$c_b: [\mu(c_b) - \sigma(c_b),\ \mu(c_b) + \sigma(c_b)] \to [0,1]$

$c_m(x,y) = median\{c_b(x+a,\ y+b) \mid -r \le a \le r,\ -r \le b \le r\}$

$\boldsymbol{P}(i) = p_i: [\min(c_m), \max(c_m)] \to [0,1]$

**Output:** $\boldsymbol{P}$, the processed image stack

Within Algorithm 1, the first task is to generate the background image stack $\boldsymbol{B}(i,k)$. This task is performed by taking the local mean, represented by the summation of frames divided by the full temporal window, where the $i$-th frame is time-centered For each frame in the stack, the corresponding background frame is subtracted from the original image. The difference frame $d$ is then masked by the corresponding $i$-th mask that centers the laser source based on the toolpath, resulting in the cropped frame $c$. A low-pass Butterworth filter is used to improve the signal-to-noise ratio. The image $c_b$ is then contrast-adjusted to an intensity range limited to one standard deviation above and below the mean value of $c_b$. A median filter is used to further smooth local intensity variation while preserving the stark phase-contrast boundary around the keyhole. To ensure consistency of frame-to-frame intensities, the final processed image $p_i$ is normalized. **Figure 2** shows a cropped image of the raw x-ray and the preprocessed ones.

a) 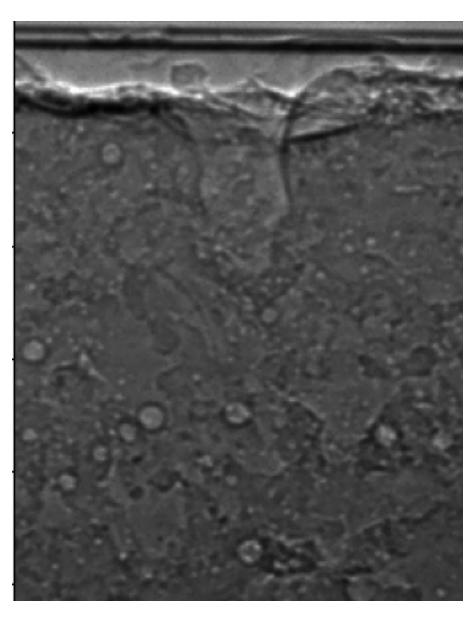 b) 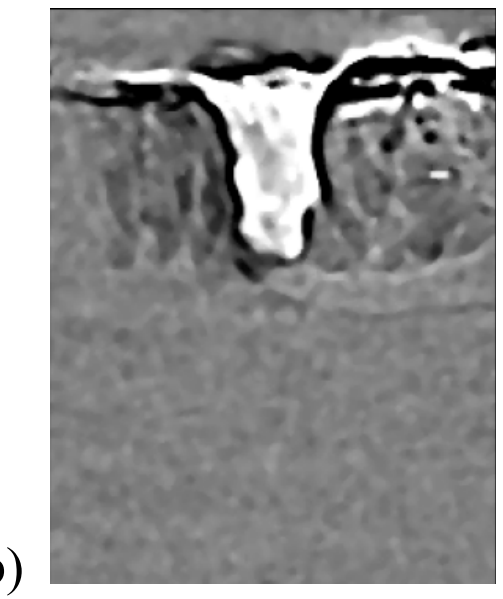

**FIGURE 2:** (a) Shows the raw x-ray image, (b) Shows the output of the preprocessor, highlighting enhanced contrast of the keyhole.

### 2.2 Segmentation Algorithm

The contrast improvements from the preprocessing simplifies the segmentation process, but additional care must be taken to properly account for temporal changes in the keyhole. While the keyhole appears bright in the image, so do other objects with relative motion, such as powder, spatter, and porosity not yet captured by the solidification front. Algorithm 2 takes the preprocessed images $\boldsymbol{P}$ obtained from Algorithm 1 and outputs the segmented keyhole image stack $\boldsymbol{S}$. Symbols employed in Algorithm 2 include $n$ and $L$, which denote the multi-level Otsu threshold count and the number of histogram bin parameters, respectively; $A_p$ and $A_h$, the area thresholds used to remove isolated pixel regions and fill holes; $D_o$ and $D_d$, the morphological opening and dilation kernels; and $\tau$, the desired core region IoU (Intersection over Union) threshold.

**Algorithm 2:** Segmentation of Images

**Input:** $\boldsymbol{P},\ n,\ L,\ A_p,\ A_h,\ D_o,\ D_d,\ \tau$

$\tilde{P} = median(\boldsymbol{P})$

$\tilde{\boldsymbol{k}}_n^* = \arg \max_{0 \le k_1 < \cdots < k_n < L} \left[\sigma_B^2(k_1, k_2, \dots, k_n,\ \tilde{P})\right]$

**for** $i = 1,2,\dots,N$ **do**

$\boldsymbol{k}_{i,n}^* = \arg \max_{0 \le k_1 < \cdots < k_n < L} [\sigma_B^2(k_1, k_2, \dots, k_n, p_i)]$

$I_i = \left(p_i > \boldsymbol{k}_{i,n-1}^* \cup \tilde{p} > \tilde{\boldsymbol{k}}_{n-1}^*\right) \cap \left(p_i > \boldsymbol{k}_{i,1}^*\right)$

$C_i = \left(p_i > \boldsymbol{k}_{i,n-1}^* \cap \tilde{p} > \tilde{\boldsymbol{k}}_{n-1}^*\right)$

$T_i = \mathcal{H}\left(\mathcal{P}\left(I_i, A_p\right), A_h\right)$

$L_i = Label\left((T_i \ominus D_o) \oplus D_o\right)$

$A_L(j) = \sum \delta_{L_i,j},\ A_C = \sum C_i,\ A_\cap(j) = \sum\left(\delta_{L_i,j} \cap C_i\right)$

$IoU(j) = \frac{A_\cap(j)}{A_L(j) + A_C - A_\cap(j)}$

$\mathcal{J}_{valid} = \{j \in \mathcal{J} \mid IoU(j) \ge \tau\}$

$V_i = \sum_{j \in \mathcal{J}_{valid}} \delta_{L_i,j}$

$\boldsymbol{S}(i) = S_i = T_i \cap (V_i \oplus D_d)$

**Output:** $\boldsymbol{S}$, the segmented binary image stack

The multi-level Otsu method [26] is performed on the median and $i$-th image frames to limit the decision space of MOSAIC for distinguishing intensity bands, while still being adaptable to the imaging conditions. $I_i$ is the inclusive mask that MOSAIC uses for initially deciding the regions of importance. A cleaned test mask is generated, $T_i$ , by using morphological operations $\mathcal{P}$ and $\mathcal{H}$ to reduce salt-and-pepper noise. Before labelling, the image is morphologically opened by kernel $D_o$ to separate the cleaned regions. The IoU of the label image and the core mask $C_i$ is found by vectorizing the label image using the Kronecker delta and determining the areas $A_L(j)$ and $A_\cap(j)$ for each $j$-th label present in $L_i$. Only unique labels, $\mathcal{J}_{valid}$, were kept out of the total $\mathcal{J}$ set based on whether the IoU value is above or below the threshold set by $\tau$. A value of $\tau < 0.1$ was used for all testing, as this ensured the core region is included while still excluding other bright regions such as porosity. A high threshold value should be avoided as the core region is, by definition, a smaller approximation of the keyhole. The valid mask $V_i$ is then generated by summing only the contributions from the valid labels. $V_i$ is morphologically dilated to reconnect regions that are considered valid, reuniting continuous regions broken up by the earlier opening operation. Finally, the segmented mask, $S_i$ is given by the union of the dilated $V_i$ mask and the original test mask $T_i$. A sample diagram is included below to show the segmentation steps.

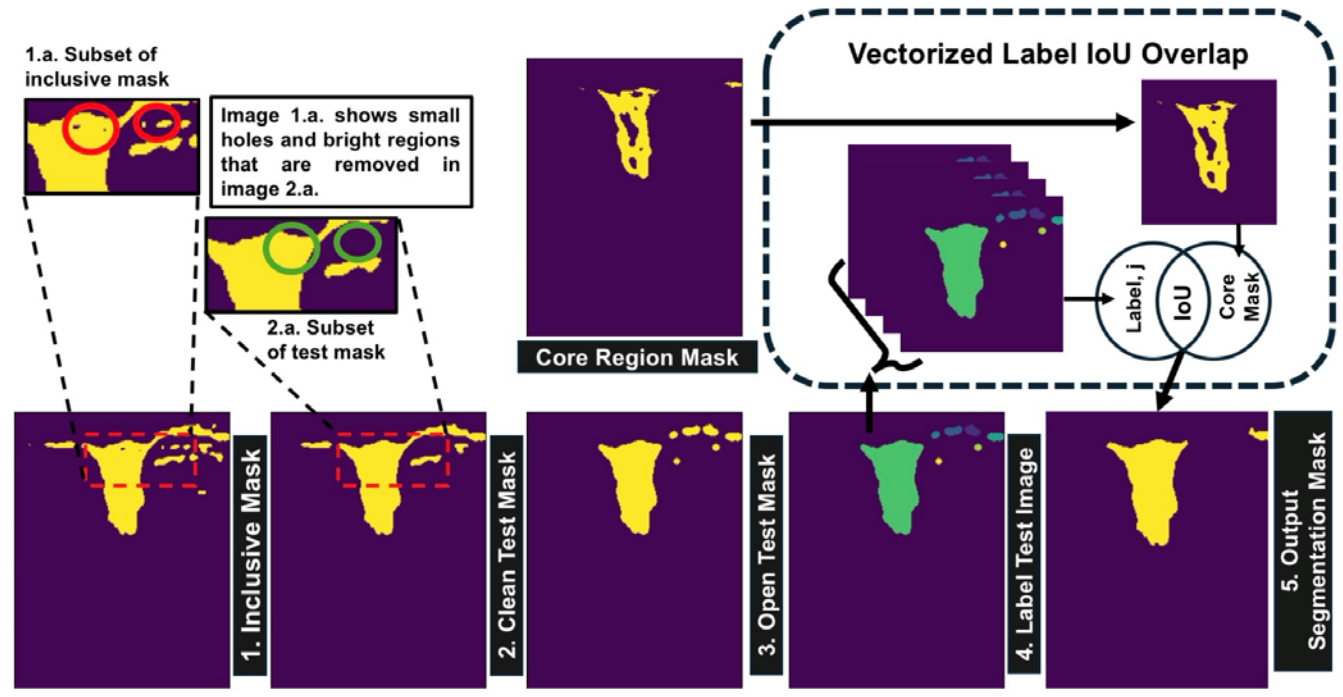

**FIGURE 3:** Diagram showing how each mask improves the segmentation output by selectively removing or adding pixels to the final segmentation output.

### 2.3 Validation and Timing Study

Validation of our results for the 12 test samples consisted of comparing the produced segmentation with manually labeled keyhole annotations. Manual labeling was aggregated from three separate labels generated by three independent annotators, giving 9 labels for each frame. The validation label was generated using pixel-wise majority voting across the 9 annotated frames, with pixels retained in the final label when annotated in more than half of the frames. For each experiment, 10% of the frames segmented were compared with the frames being selected at random from the total frame list; this validation process was selected due to its similarity to that used in the evaluation of machine learning models. The validation metrics used were the IoU, F1 score, accuracy, and precision, represented by Equations 1-4.

$$\text{IoU} = \frac{\Sigma(K \cap T)}{\Sigma(K \cup T)} \tag{1}$$

$$\text{F1} = 2 * \frac{\text{IoU}}{1 - \text{IoU}} \tag{2}$$

$$A = \frac{1}{hw}\Sigma(K = T) \tag{3}$$

$$P = \frac{\Sigma K \cap T}{\Sigma K} \tag{4}$$

To understand how MOSAIC reacts to changes in the input parameters, tests were performed by sweeping the value of selected parameters and observing the impacts on the F1 score. The parameters that will be tested for our sensitivity study are the input values for the substrate cutoff, subset within the toolpath mask $\boldsymbol{M}$, and the temporal window size, $k$, for averaging the background as they heavily influence segmentation performance. We also investigate the impact of the median filter kernel size $r$ on both the timing of the preprocessor and the average F1 score. The median filter helps to smooth the image, but at the cost of significant computational time, so optimization is critical.

In order to justify the use of MOSAIC over other commonly available machine-learning-based segmentation models, MOSAIC was compared against a trained You Only Look Once (YOLO) model and a Segment Anything Model (SAM) [21,24]. The YOLO model was fine-tuned using a published YOLOv26-seg backbone model and then given 242 and 69 labeled images to train and validate on, respectively, from Samples A, F, G, and L. These samples were chosen to expose the YOLO to all major categories of samples while having a test set to compare the YOLO against with distinct features not seen in training. Testing of the YOLO model was performed on all samples. The SAM version 2.1 base model was tested using two approaches. The first test of SAM was with a human-in-the-loop, utilizing a GUI to select positive and negative seed points for the model to initiate from. The second test was to compare the automated performance of SAM, using a seed point determined by the median experiment image for all images in the set. Beyond the automated performance of SAM, the second SAM test was used for timing the inference of the model. All YOLO and SAM tests were given background-subtracted images (***P***) without further processing.

Timing was performed using the built-in timer function in Python. The total time is reported as the amount of CPU time required per image processed by MOSAIC. As memory read and write speeds can vary widely based on device, the image loading time was not considered in our timing study, only the time to perform the processing and segmentation operations. To demonstrate the low computational overhead needed to run MOSAIC, a laptop was used for timing studies. The laptop has an Intel Core i7-10750H CPU, a dedicated NVIDIA Quadro T1000 Max-Q with 4GB of VRAM, 16 GB of RAM, and a 512

GB NVMe SSD, running Windows 11. For the machine learning models, the CPU and CUDA 12.4 GPU performance was recorded, with the inference time per image averaged across the tested images.

## 3. RESULTS AND DISCUSSION

Through the developed algorithm MOSAIC, we can clearly segment the keyhole from the background regardless of whether the underlying substrate is clean or obscured. In this section, we will discuss the image output qualitatively (Section 3.1), the validity of the results as compared to machine-learning methods (Section 3.2), and the sensitivities of the segmented results to the parameters in our algorithms (Section 3.3). Timing metrics (Section 3.4) are also provided, demonstrating the rapid nature of MOSAIC on low-end computer resources.

### 3.1 Image Output from MOSAIC

The results shown in **Figure 4** highlight the clear segmentation boundaries that are produced by MOSAIC. The algorithm can make distinctions between not just the keyhole and the substrate, but also between the keyhole and any powder on the bed, powder trapped along the sides of the substrate, or porosity that might obscure the view of the keyhole from a brief glance.

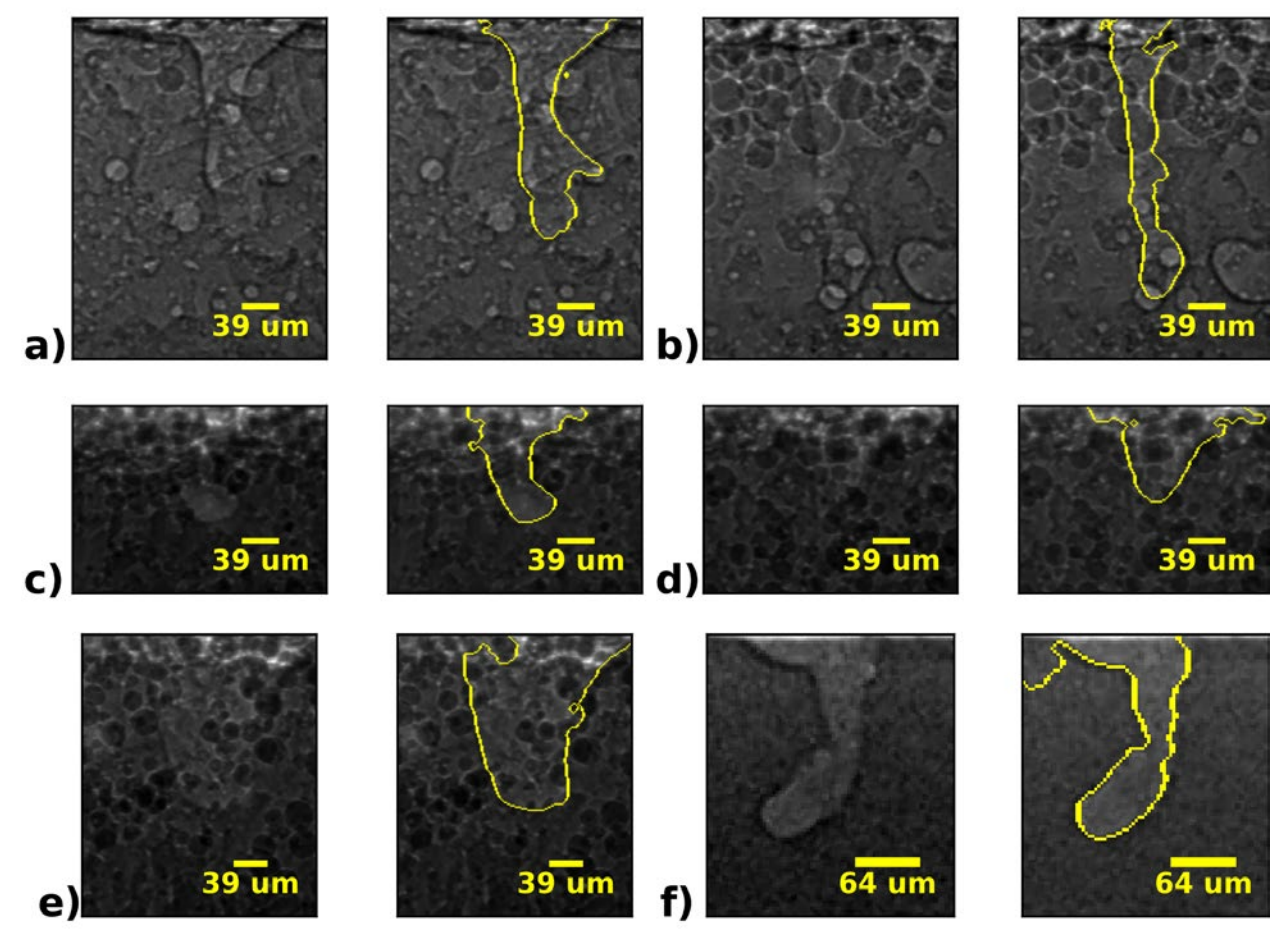

**FIGURE 4**: Raw and segmented images using MOSAIC for single frames from samples E, F, G, C, D, and K in order of a)-f). a) and b) show high energy input into an aluminum substrate with and without powder respectively. c) and d) show reduced contrast titanium samples with powder obstructing the keyhole. e) shows a wide titanium keyhole. f) highlights a 1 MHz sample.

**Figures 4b-4e** show powder particles that have obscured the region just below the top surface of the substrate. Regardless, MOSAIC assembles the keyhole structure deterministically to successfully capture the important morphological features of the gas-metal interface at the melt pool surface. In **Figure 4b**, dense regions of porosity and powder are trapped between the substrate and the glassy carbon used to hold the powder bed. As these obstructions remain relatively motionless, they are removed from the frame by time-averaged background subtraction. While obstructions may limit the contrast or brightness across portions of the keyhole region, the preprocessing steps can evidently cut through to provide accurate and consistent segmentation. **Figures 4a** and **4f** show cases where there is no powder on the substrate. Again, Figure 4a has porosity within the as-built substrate, which is correctly ignored by MOSAIC's segmentation. **Figure 4f** demonstrates MOSAIC's ability to maintain consistent segmentation when motion from frame to frame is minimal, as **4f** is from the 1 MHz experiments, while other images are from the experiments using the 50 kHz sampling rate.

### 3.2 Validation Study

For all validated samples, regardless of variations in material, powder presence, sampling rates, or evaluation criteria, MOSAIC demonstrates little deviation (< 3%) from the average of all sample validation metrics, as shown in **Figure 5**.

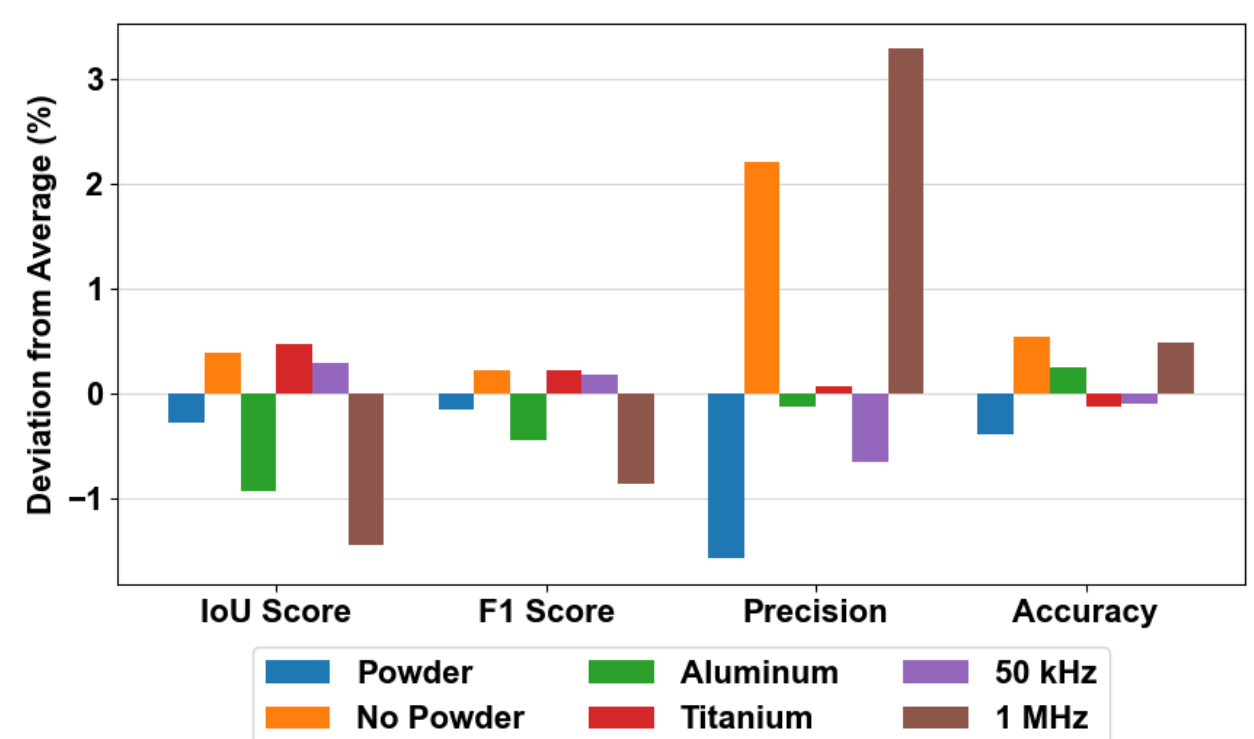

**FIGURE 5:** Deviation from the average validation metric for different subsets of samples. The powder/no-powder samples have near zero deviation from the average in IoU and F1.

For clarification, the average value refers to the average value across all samples, while the mean value score refers to a specific subset of samples. **Table 2** gives the average IoU, F1 score, accuracy, and precision across all samples. MOSAIC demonstrated an average across all samples of a ~90% F1-score and a >95% precision. Qualitative observations showed that the segmentation result fell 2-3 pixels inside the boundaries of the manually segmented frames; this is likely due to the morphological opening operation, because information is removed from the control image before the algorithm makes a final decision on the keyhole morphology. Compared to the ML models tested, MOSAIC performs equal to or better than all non-human-prompted models, with only the manual SAM segmentation performing slightly better than MOSAIC. This is expected, as the human-prompted SAM segmentation has the ability to correct large errors through negative prompt points. The YOLO model performs approximately equal to MOSAIC on the training samples, but performance reduces significantly for the unseen test samples. This indicates that the YOLO model used here struggles to generalize effectively across datasets with alternate imaging conditions or keyhole morphologies. In contrast, MOSAIC is able to handle a wide range of image quality and keyhole geometries without any major alteration of parameters used in the model. Table 2 highlights this disparity, as while both MOSAIC and YOLO-training datasets have a ~0.05 difference in F1 scores between their average and worst performances, the YOLO-test dataset has a more significant ~0.30 difference in F1 scores. The human-in-the-loop SAM tests show comparable performance, but the automated implementation decreases in performance. This shows that the SAM relies heavily on manual prompting to properly segment objects, limiting applicability to automated analysis methods.

**TABLE 2:** Average performance across the dataset tested and worst performing samples for MOSAIC as well as other machine-learning models, 1 represents a perfect score.

| Average Performance | | | | |
|---|---|---|---|---|
| Model | IoU | F1 | Precision | Accuracy |
| MOSAIC | 0.811 | 0.894 | 0.953 | 0.981 |
| YOLO (training) | 0.806 | 0.890 | 0.929 | 0.987 |
| YOLO (testing) | 0.785 | 0.867 | 0.926 | 0.975 |
| SAM 2.1 (manual) | 0.825 | 0.902 | 0.906 | 0.984 |
| SAM 2.1 (auto) | 0.770 | 0.853 | 0.863 | 0.966 |
| Worst Performing Sample Scores | | | | |
| Model | IoU | F1 | Precision | Accuracy |
| MOSAIC | 0.738 | 0.846 | 0.881 | 0.933 |
| YOLO (training) | 0.712 | 0.830 | 0.886 | 0.982 |
| YOLO (testing) | 0.442 | 0.591 | 0.800 | 0.860 |
| SAM 2.1 (manual) | 0.735 | 0.846 | 0.869 | 0.947 |
| SAM 2.1 (auto) | 0.663 | 0.759 | 0.743 | 0.921 |

A more detailed look at the deviations from the average in **Figure 5** shows interesting trends between the different groups of samples. For the powder, no powder, 50 kHz, and 1 MHz samples, the direction of deviation from the average is the same for IoU and F1, but it flips for the accuracy and precision validation metrics. The increased accuracy and precision and decreased IoU and F1 scores of the no powder and 1MHz cases support the conclusion that MOSAIC is more likely to under-predict the size of the keyhole than to over-predict. Under-prediction is more pronounced in the no powder and 1 MHz cases, as they have less frame-to-frame variation and MOSAIC is programmed to interpret low-value pixels in the median image as background pixels. Over-prediction is not as pronounced for the powder and 50 kHz samples, with the performance difference being attributed to the powder obstruction causing a reduction of the dynamic range. Looking at the aluminum and titanium alloy samples, all aluminum samples performed better than average, and all titanium samples performed worse than average. The difference in the x-ray attenuation caused by the material selected and its thickness impacts the dynamic range, ultimately reducing the contrast which slightly decreases the segmentation performance.

### 3.3 Parameter Sensitivity Study

The interpretable parameters of the MOSAIC algorithm can help users select the proper input settings. In the first parameter study, we analyzed how the F1 score varied with the vertical crop used across different samples. **Figure 6** clearly shows the importance of proper substrate cropping for maintaining a high F1 score and its dependence on the substrate material's x-ray attenuation. Aluminum substrates tested, with or without powder, are not impacted significantly by the determination of substrate cutoff. Once the cutoff goes beneath the substrate the F1 score varies much more in all samples, as MOSAIC has less information to properly determine the behavior at the melt pool surface. The titanium samples show a significantly worse F1 score when the area above the substrate is included. The difference in intensities between the bright incident beam and the dim substrate region causes the preprocessor to effectively output a binary image, which results in the keyhole sitting in a gray-value region with insufficient contrast to be properly segmented. This behavior informs the user that regions with high contrast boundaries should be cropped out to ensure proper segmentation.

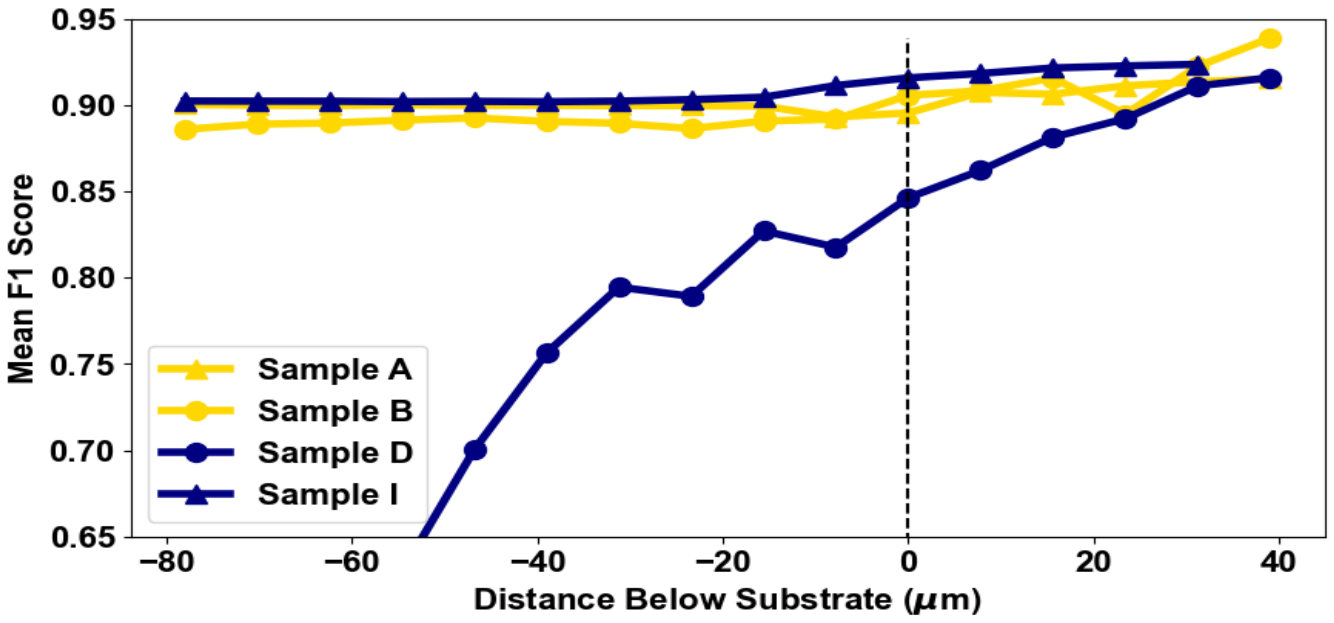

**FIGURE 6:** Impact of the input vertical crop of the frame and on the samples F1 score, comparing samples of different material x-ray attenuation.

The temporal window size, $k$, is important to ensure the highlighting of the moving keyhole through proper background subtraction. Increasing the temporal window can make the moving object easier to segment by improving contrast between the object and background. However, the temporal window must not exceed the number of frames in the experimental stack, and smaller windows may be preferred to preserve highly time-resolved variation instead of maintaining high keyhole contrast. From our window size parameter study, a normalized temporal window metric can be determined to compare the F1 score of samples with different scan speeds and frame rates. The normalized temporal window, $\bar{k}$ is defined in Equation 5,

$$\bar{k} = \frac{kv}{\omega \Delta x} \tag{5}$$

where $k$ is the previously mentioned window size, $v$ is the moving reference frame speed in meters per second, $\omega$ is the frame frequency in Hz, and $\Delta x$ is the pixel length scale in meters. **Figure 7** shows a plateauing behavior, with the different samples tested approaching an optimal value as the normalized temporal window, $\bar{k}$, approaches 50. This cutoff represents the number of pixels it takes for the object to pass, as $v$ divided by $\omega \Delta x$ represents the pixel distance between frames. The normalized temporal window metric can assist users of MOSAIC to pre-determine a minimum temporal window size to ensure proper segmentation. Reliability of the normalized temporal window across samples shows the deterministic and predictable nature of the segmentation algorithm.

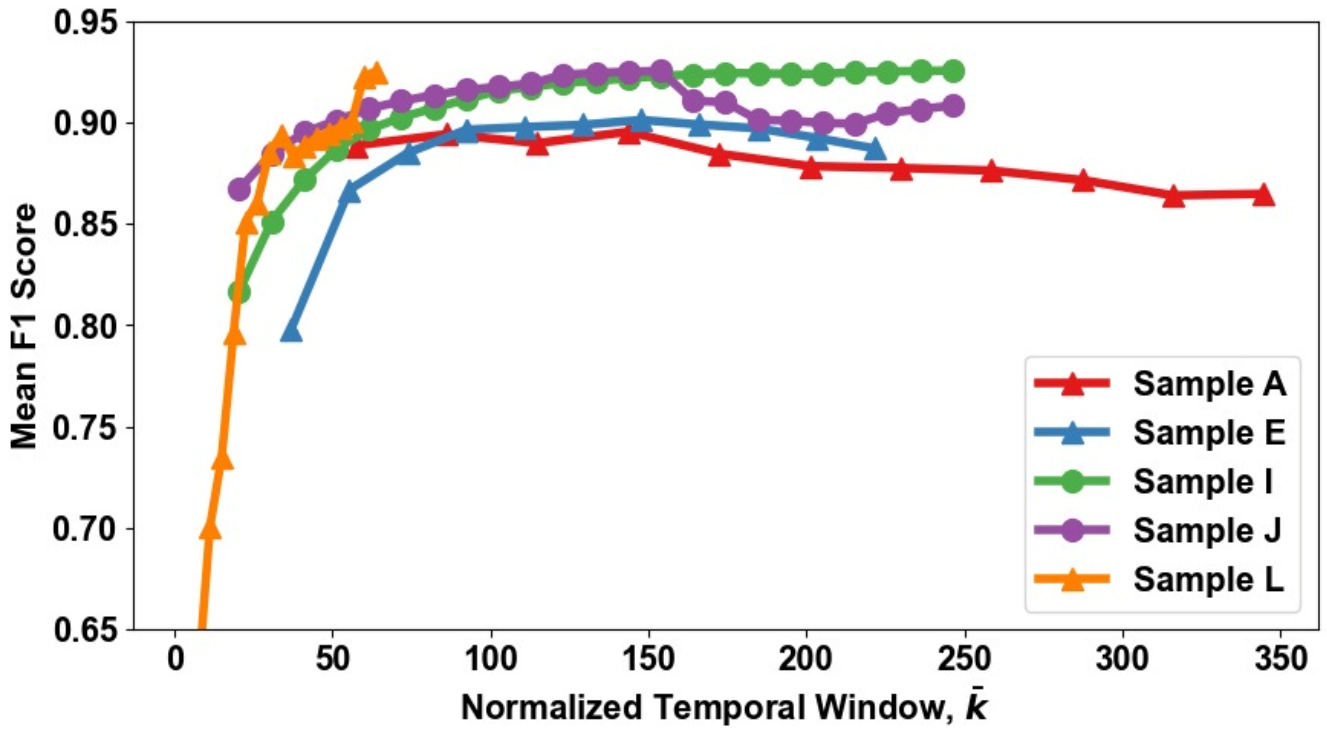

**FIGURE 7:** Normalized temporal window versus the F1 score of samples A, E, I, J, and L, corresponding to object speeds of 1.4, 0.9, 0.5, 0.5, and 0.6 m/s, respectively.

## 3.4 Timing Study

Having now validated MOSAIC's ability to properly segment the keyhole, we now test the efficiency of the algorithm and compare it to commonly available ML methods. The ideal tool is one that can give the correct result instantly, enabling rapid on-site usage. **Table 3** demonstrates that we can preprocess and segment images at a rate of 19.90 milliseconds per image for our large image samples (150 x 250 pixels) and 10.52 milliseconds per image for the smaller image samples (80 x 190 pixels). For ML models, all images were combined as these methods require normalization to a common image size, typically a number divisible by 32. For the YOLO model, a square image size of 320 was used, as this would prevent down-sampling from occurring on larger images. For the SAM 2.1 base model, a square image size of 1024 was used, as smaller image sizes were not able to be properly inferred by the fine-tuned model. Across all processor types and machine learning models, MOSAIC outperformed the two ML models in speed. The improved performance of MOSAIC stems from eliminating the explicit inference step typically required for image segmentation. All machine learning models tested consist of large matrices that need to be inferred, which inherently limit the ability to speed up the performance. Exacerbating the issue, attempts to improve the generalizability of ML models can increase model complexity and inference cost, further reducing computational efficiency. MOSAIC does not require ML inference and therefore avoids this limitation.

As high-speed operando x-ray imaging produces increasingly large data volumes, the per-frame speed difference becomes practically important for on-site analysis, where users may not have access to high-end GPU hardware. Using the per-image processing times reported in Table 3, a representative 15,000-frame stack would require approximately 5.0 min with MOSAIC on CPU, compared with 13.5 min for YOLO on CPU and 22.0 h for SAM on CPU. This runtime advantage is compounded by the fact that ML methods require training or fine-tuning datasets, which can require substantial manual annotation time before deployment; for example, 1,000 images labeled by three annotators with three repeated annotations per image would require approximately 12.5-25 h of annotation time, assuming 5-10 s per labeled mask.

**TABLE 3:** Results of timing study between large and small image samples for MOSAIC, and a comparison between CPU and GPU-based timing for common AI segmentation models

| Model | Processor | Time (ms/img) | |
|---|---|---|---|
| | | Large Image | Small Image |
| MOSAIC - Total | CPU | 19.90 $\pm$ 2.67 | 10.19 $\pm$ 0.191 |
| -Preprocessing | | 12.24 $\pm$ 1.76 | 5.87 $\pm$ 0.162 |
| -Segmentation | | 7.76 $\pm$ 1.03 | 4.31 $\pm$ 0.039 |
| YOLO | CPU | 54.06 $\pm$ 15.41 | |
| | GPU | 32.63 $\pm$ 5.64 | |
| SAM 2.1 | CPU | 5284 $\pm$ 269 | |
| | GPU | 884.2 $\pm$ 82.2 | |

**Figure 8** plots the F1 score and the median filter computation time of samples L, G, and A against an increase in the pixel size of the median filter kernel. As mentioned in the methods section, the median filter works as an edge-preserving filter to reduce noise that may come from small powder movement, pore motion, or substrate movement due to the thermal heating effect. At low kernel dimensions, the F1 score of all samples is relatively similar, with a slight improvement in performance at a kernel size of ~4 pixels. As the kernel size increases, computation time increases similarly, as the median filter must sort a larger number of pixels to determine the median value of the region. The F1 score begins to decrease at high kernel sizes as smaller features within the image are degraded or disappear entirely. Therefore, to ensure both the validity of the segmentation and reasonable computation times, a low kernel size between 1 and 5 pixels in radius should be used in the median filter to smooth out noise while maintaining good temporal performance. The previous images shown in **Figure 4.** and for the results given in **Table 2.**, a kernel radius size of 3 pixels was used.

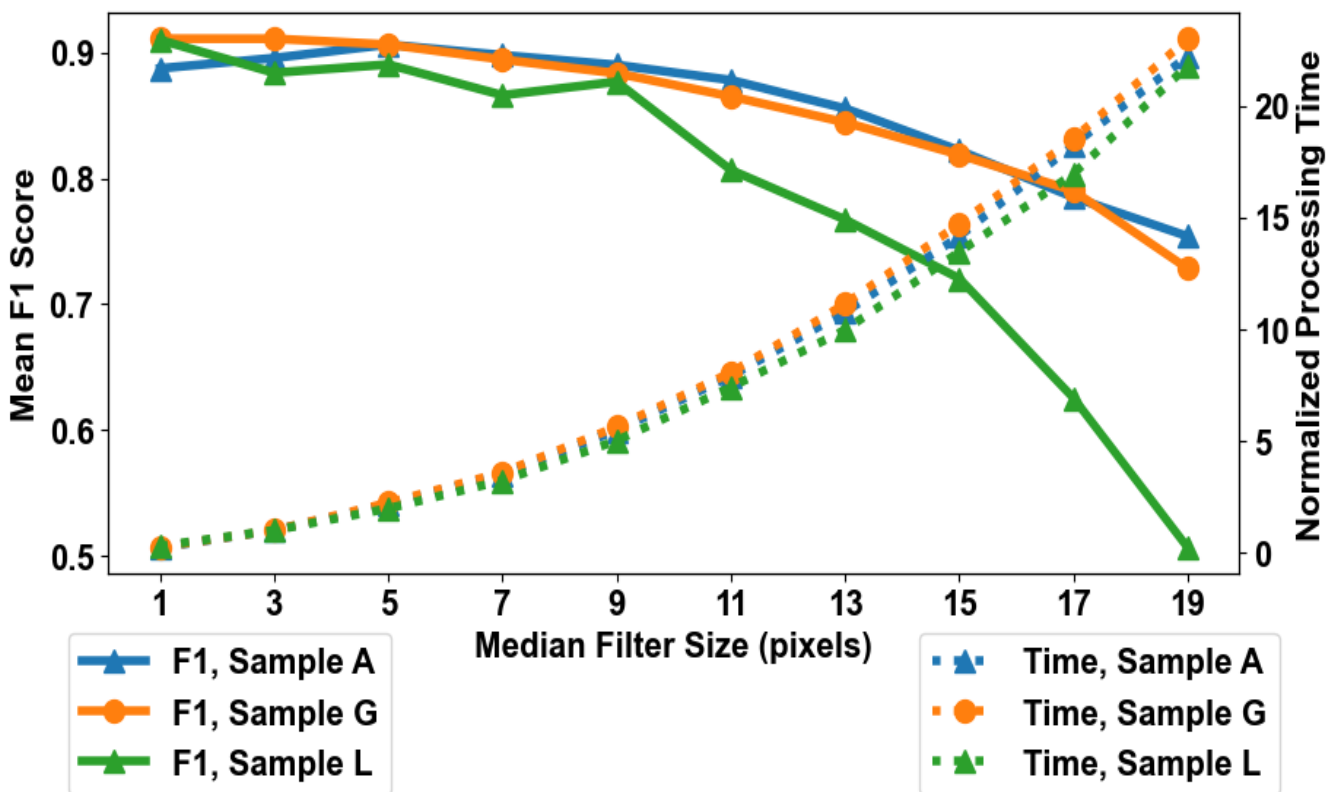

**FIGURE 8:** Plot showing the impact of changing median filter kernel size on F1 score and computation time during preprocessing.

## 4. CONCLUSION

In this work, MOSAIC was developed and validated as a first-principles-based framework for automated analysis of LPBF synchrotron x-ray imaging data. By eliminating the need for training or fine-tuning datasets, the framework enables rapid characterization of keyhole behavior during high-speed operando experiments while maintaining strong segmentation performance across diverse experimental conditions.

- Demonstrated accurate and validated segmentation of LPBF keyhole morphology across diverse experimental conditions.
- Achieved superior computational efficiency and robustness compared to conventional machine learning-based segmentation approaches.
- Provides a foundation for real-time characterization of melt pool behavior during operando x-ray experiments, supporting advanced process monitoring and control.

Future iterations of MOSAIC will include modifications to allow for an expanded set of analysis tools for segmenting the melt pool boundary, powder particles, and pore motion.

## ACKNOWLEDGEMENTS

The authors would like to acknowledge the support received from the DEVCOM Army Research Laboratory under cooperative agreement W911NF2120199 and W911NF2520035 (JC and GM), and NSF Grant DGE-2234667 (CT). The views and conclusions contained in this document are those of the authors and should not be interpreted as representing official policies, either expressed or implied, of the Army Research Laboratory or the US Government. The US Government is authorized to reproduce and distribute reprints for Government purposes, notwithstanding any copyright notation herein. The data was generated using the DOE Advanced Photon Source (DE- AC02–06CH11357).

The authors would like to acknowledge Samuel Clark and Kamel Fezzaa at Argonne National Laboratory for their support of the x-ray imaging experiments used in this paper.

The authors would also like to acknowledge the feedback and support from Prof. Tao Sun and Dr. Jon-Erik Mogonye during the development of this work.

## REFERENCES

[1] Liao S, Webster S, Huang D, Council R, Ehmann K, Cao J. Simulation-guided variable laser power design for melt pool depth control in directed energy deposition. Additive Manufacturing 2022;56:102912. https://doi.org/10.1016/j.addma.2022.102912.

[2] Olleak A, Dugast F, Bharadwaj P, Strayer S, Hinnebusch S, Narra S, et al. Enabling Part-Scale Scanwise process simulation for predicting melt pool variation in LPBF by combining GPU-based Matrix-free FEM and adaptive Remeshing. Additive Manufacturing Letters 2022;3:100051. https://doi.org/10.1016/j.addlet.2022.100051.

[3] Hasan N, Saha AK, Wessman A, Shafae M. Machine learning-based layer-wise detection of overheating anomaly in LPBF using photodiode data. Manufacturing Letters 2024;41:1423–31. https://doi.org/10.1016/j.mfglet.2024.09.169.

[4] Carter FM, Porter C, Kozjek D, Shimoyoshi K, Fujishima M, Irino N, et al. Machine learning guided adaptive laser power control in selective laser melting for pore reduction. CIRP Annals 2024;73:149–52. https://doi.org/10.1016/j.cirp.2024.04.043.

[5] Kozjek D, Carter FM, Porter C, Mogonye J-E, Ehmann K, Cao J. Data-driven prediction of next-layer melt pool temperatures in laser powder bed fusion based on co-axial high-resolution Planck thermometry measurements. Journal of Manufacturing Processes 2022;79:81–90. https://doi.org/10.1016/j.jmapro.2022.04.033.

[6] Yan W, Lin S, Kafka OL, Lian Y, Yu C, Liu Z, et al. Data-driven multi-scale multi-physics models to derive process–structure–property relationships for additive manufacturing. Comput Mech 2018;61:521–41. https://doi.org/10.1007/s00466-018-1539-z.

[7] Raj A, Huang D, Stegman B, Abdel-Khalik H, Zhang X, Sutherland JW. Modeling spatial variations in co-axial melt pool monitoring signals in laser powder bed fusion. Journal of Manufacturing Processes 2023;89:24–38. https://doi.org/10.1016/j.jmapro.2022.12.048.

[8] Yeung H. Toward real-time feedback control for powder bed fusion additive manufacturing. Additive Manufacturing 2025;114:105041. https://doi.org/10.1016/j.addma.2025.105041.

[9] Zhao C, Fezzaa K, Cunningham RW, Wen H, De Carlo F, Chen L, et al. Real-time monitoring of laser powder bed fusion process using high-speed X-ray imaging and diffraction. Sci Rep 2017;7:3602. https://doi.org/10.1038/s41598-017-03761-2.

[10] Parab ND, Zhao C, Cunningham R, Escano LI, Fezzaa K, Everhart W, et al. Ultrafast X-ray imaging of laser–metal additive manufacturing processes. J Synchrotron Radiat 2018;25:1467–77. https://doi.org/10.1107/S1600577518009554.

[11] Gan Z, Kafka OL, Parab N, Zhao C, Fang L, Heinonen O, et al. Universal scaling laws of keyhole stability and porosity in 3D printing of metals. Nat Commun 2021;12:2379. https://doi.org/10.1038/s41467-021-22704-0.

[12] Zhao C, Parab ND, Li X, Fezzaa K, Tan W, Rollett AD, et al. Critical instability at moving keyhole tip generates porosity in laser melting. Science 2020;370:1080–6. https://doi.org/10.1126/science.abd1587.

[13] Li X, Zhao C, Sun T, Tan W. Revealing transient powder-gas interaction in laser powder bed fusion process through multi-physics modeling and high-speed synchrotron x-ray imaging. Additive Manufacturing 2020;35:101362. https://doi.org/10.1016/j.addma.2020.101362.

[14] Bitharas I, Parab N, Zhao C, Sun T, Rollett AD, Moore AJ. The interplay between vapour, liquid, and solid phases in laser powder bed fusion. Nat Commun 2022;13:2959. https://doi.org/10.1038/s41467-022-30667-z.

[15] Guo D, Lambert-Garcia R, Hocine S, Fan X, Greenhalgh H, Shahani R, et al. Correlative spatter and vapour depression dynamics during laser powder bed fusion of an Al-Fe-Zr alloy. Int J Extrem Manuf 2024;6:055601. https://doi.org/10.1088/2631-7990/ad4e1d.

[16] Ruckh E, Lambert-Garcia R, Hocine S, Getley ACM, Iantaffi C, Bhatt A, et al. Process mapping of Ti-6Al-4V laser powder bed fusion using in situ high-speed synchrotron x-ray imaging. Additive Manufacturing 2025;113:105007. https://doi.org/10.1016/j.addma.2025.105007.

[17] Huang Y, Fleming TG, Clark SJ, Marussi S, Fezzaa K, Thiyagalingam J, et al. Keyhole fluctuation and pore formation mechanisms during laser powder bed fusion additive manufacturing. Nat Commun 2022;13:1170. https://doi.org/10.1038/s41467-022-28694-x.

[18] Hojjatzadeh SMH, Parab ND, Guo Q, Qu M, Xiong L, Zhao C, et al. Direct observation of pore formation mechanisms during LPBF additive manufacturing process and high energy density laser welding. International Journal of Machine Tools and Manufacture 2020;153:103555. https://doi.org/10.1016/j.ijmachtools.2020.103555.

[19] Carter FM, Kozjek D, Porter C, Clark SJ, Fezzaa K, Fujishima M, et al. Melt pool instability detection using coaxial photodiode system validated by in-situ X-ray imaging. CIRP Annals 2023;72:205–8. https://doi.org/10.1016/j.cirp.2023.03.031.

[20] Samaei A, Leonor JP, Gan Z, Sang Z, Xie X, Simonds BJ, et al. Benchmark study of melt pool and keyhole dynamics, laser absorptance, and porosity in additive manufacturing of Ti-6Al-4V. Prog Addit Manuf 2025;10:491–515. https://doi.org/10.1007/s40964-024-00637-6.

[21] Ravi N, Gabeur V, Hu Y-T, Hu R, Ryali C, Ma T, et al. SAM 2: Segment Anything in Images and Videos 2024. https://doi.org/10.48550/ARXIV.2408.00714.

[22] Virtanen P, Gommers R, Oliphant TE, Haberland M, Reddy T, Cournapeau D, et al. SciPy 1.0: fundamental algorithms for scientific computing in Python. Nat Methods 2020;17:261–72. https://doi.org/10.1038/s41592-019-0686-2.

[23] van der Walt S, Schönberger JL, Nunez-Iglesias J, Boulogne F, Warner JD, Yager N, et al. scikit-image: image processing in Python. PeerJ 2014;2:e453. https://doi.org/10.7717/peerj.453.

[24] Jocher G, Qiu J, Liu M, Lyu S, Akyon FC, Kalfaoglu ME. Ultralytics YOLO26: Unified Real-Time End-to-End Vision Models 2026. https://doi.org/10.48550/ARXIV.2606.03748.

[25] Ren Z, Gao L, Clark SJ, Fezzaa K, Shevchenko P, Choi A, et al. Machine learning–aided real-time detection of keyhole pore generation in laser powder bed fusion. Science 2023;379:89–94. https://doi.org/10.1126/science.add4667.

[26] Liao P-S, Chen T-S, Chung P-C. A Fast Algorithm for Multilevel Thresholding. Journal of Information Science and Engineering 2001;17:713–27. https://doi.org/10.1688/JISE.2001.17.5.1.